\documentstyle[aps,multicol]{revtex}

\input{tcilatex}

\begin{document}
\title{High frequency magneto-impedance of double perovskite $%
La_{1.2}Sr_{1.8}Mn_{2}O_{7}$: secondary transitions at high temperatures}
\author{P. V. Patanjali, P. Theule, Z. Zhai, N. Hakim, and S. Sridhar$^{a}$}
\address{Physics Department, Northeastern University, 360 Huntington Avenue, Boston,\\
MA 02115}
\author{R. Suryanarayanan, M. Apostu, G. Dhalenne, and A. Revcolevschi,}
\address{Laboratoire de Chimie des Solides, UA 446, Universit\v{e} Paris-Sud, 91405\\
Orsay, France}
\maketitle

\begin{abstract}
Radio frequency magneto-impedance measurements reveal a pronounced anomaly
at $260K$ besides the main MI transition at $125K$ in the double perovskite
material $La_{1.2}Sr_{1.8}Mn_{2}O_{7}$. This feature is not seen clearly in
static resistivity and magnetization measurements. We suggest that this
anomaly represents short-range magnetic correlations enhanced at radio
frequencies, with the easy axis along the $c-$axis.
\end{abstract}

\smallskip 
\begin{multicols}{2}%

The observation of colossal magnetoresistance $(CMR)$ in perovskite
manganites $La_{1-x}A_{x}MnO_{3}$ $(A=Sr,Ca)$ $[113]$ has triggered further
research into related systems\cite{von,jin,cnrrao,ramirez}. The CMR in these
materials is normally observed at temperatures close to the
ferromagnetic-to-paramagnetic transition associated with a metal insulator $%
(MI)$ transition. Following the double exchange mechanism of Zener\cite
{zener}, based upon the strong on-site (Hund's rule) coupling between the
charge carriers (e$_{g}$-like state) and the local spins (t$_{2g}$-like
state), several models\cite{pwa,degennes,kubo,furukawa,millis} have been
proposed to account for these properties, but many of the observed features
cannot be accounted by the double exchange interaction alone. These
well-studied $113$ compounds can be considered as the $n=\infty $ members of
the Ruddlesden-Popper $(RP)$ series $(La_{1-x}Sr_{x})_{n+1}Mn_{n}O_{3n+1}$
which can be described as intergrowths between $La_{1-x}Sr_{x}MnO_{3}$
perovskite blocks and rock-salt-type layers $(La,Sr)O.$

In order to examine further the relation between CMR, magnetic and MI
transitions, attention has been focussed on other materials with reduced
dimension $n$\cite{mahesh,battle,laffez,asano}. In particular, for the $n=2$
member of the RP series $(La_{1-x}Sr_{x})_{n+1}Mn_{2}O_{7}(327)$, the
material is antiferromagnetic for $x=0$ , while in the region $0.2\leq x\leq
0.4$ the material is ferromagnetic and exhibits an MI transition at $T_{c}.$
The maximum $T_{c}=125K$ obtained in this series is for $x=0.4$. Moritomo et
al. \cite{moritomo96} have reported a high value of CMR in $x=0.4$
composition at temperatures near and far away from $T_{c}$.

Most of the experiments that probe the manganites are static in nature and
have limited capabilities in elucidating the basic mechanisms involved.
Recent dynamic measurements like EPR\cite{mullerEPR} and FMR \cite{lofland}
have provided microscopic evidence for the formation of Jahn-Teller polarons
and spin wave resonance, respectively, in the manganites. In $%
La_{1-x}Sr_{x}MnO_{3}$, Srikanth et al.\cite{srikanth} have very recently
observed additional new features in the rf impedance measurements, that are
not clearly visible in the dc resistivity and magnetization measurements.
These results show that the dynamic rf measurements reveal new transitions,
and in addition also give information on the collective response of spin and
charge dynamics. In this report we describe, for the first time, temperature
and field dependence of radio frequency dynamic response of the double
perovskite $La_{1.2}Sr_{1.8}Mn_{2}O_{7}$. We focus on a pronounced anomaly
around $260K$ observed in many samples and suggest that this anomaly
represents magnetic correlations, perhaps 2-D in nature, with a single
energy and temperature scale.

Single crystals (5 cm long and 0.5 cm diameter) of $%
La_{1.2}Sr_{1.8}Mn_{2}O_{7}$ were grown from the melt by the floating-zone
method using a mirror furnace \cite{rev69}. Powder X-ray diffraction (XRD)
analysis resulted in the lattice parameters $a=3.864\pm 0.002$ \AA\ and $%
c=20.130\pm 0.006$ \AA\ , in close agreement with those previously reported 
\cite{moritomo96}. The polycrystalline material was synthesized by the
conventional solid state sintering technique. No extra phases were detected
by $XRD$ and the lattice parameters were found to be identical to those
determined in the case of the single crystals. Three single crystals labeled 
$SC1$, $SC2$ and $SC3$, and a polycyrstalline sample labeled $PC$ were
studied.

The rf measurements were carried out using an ultrastable tunnel diode
oscillator. The sample is placed in a coil which forms the inductive part of
the $LC$ resonant circuit. In this technique, a change in the effective
reactance $X$ of the sample, caused by varying temperature, $T$ magnetic
field $H$, or angle $\theta $ (between crystalographic axis and $H$) leads
to a change in the inductance of the coil which in turn results in a shift
of the resonant frequency of the oscillator. The quantity that is measured
is the change in the resonant frequency, $df(T)=[f_{0}(T_{\max })-f(T_{\max
})]-[f_{0}(T)-f(T)]$, where f$_{0}(T)$ is the empty coil frequency. From the
elementary consideration of ac impedance and Maxwell's equations for
magnetic, conducting materials, it can be shown that $df$ $=-G(dX)$ $\propto
-d(\mu \rho )^{1/2}$. Where, $dX(T)=X(T_{\min })-X(T)\ $ is the change in
reactance, $G$ the geometric factor and, $\rho (T,H)$ and $\mu (T,H)$ are
the resistivity and permeability of the magnetic material, respectively. The
change in the reactance is calculated using the above relation.

The stability of the circuit is very high, typically, $1Hz$ in $4MHz$ and
the operating frequency is in the range $2-4MHz$ depending on the coil and
sample used. The inductive coil with the sample is inserted into a Helium
flow cryostat and coupled through a rigid coaxial cable, to enable
temperature variation between $4.2K$ and $300K$. The magnetic field is
applied by placing the sample coil-cryostat-assembly between the pole pieces
of an electromagnet. The field dependent measurements were done with the
field parallel to the $a-b$ plane $(H\,||\,ab)$ and the $\hat{c}-axis$ $%
(H\,||\,c)$, and the maximum field applied is $6KOe$.

The magnetization and $CMR$ of the crystal have been described elsewhere\cite
{prellier}. Briefly, a sharp ferromagnetic to paramagnetic and metal -
insulator $\left( MI\right) \ $transition is observed around $125K$, in the
temperature dependence of magnetization, $M\left( T\right) $ and
resistivity, $\rho \left( T\right) $ measurements, respectively. However, a
strong deviation from the Curie-Weiss law was noted above $T>200K.$
Interestingly, in the case of sample $SC1$, while a very small change in
slope is observed at around $260K$ in the $M\left( T\right) $, no such
anomaly is observed in the $dc$ $\rho \left( T\right) $ in both $ab-$ $plane$
$(\rho _{ab})$ and $\hat{c}-axis$ $\left( \rho _{c}\right) $ directions
(inset, Fig. 2). From the field dependence of magnetization $M(H)$ at $5K$
it is found that the $easy-axis$ of magnetization lies in the $ab-plane$
which is in agreement with reported results\cite{kimura96,bader98}. The
magnetic moment at $H=5T$ was found to be $3.5\mu _{B}/Mn$ atom which is in
close agreement with the theoretical value, $3.6\mu _{B}/Mn$ atom, ensuring
the almost $100\%$ spin polarization of the conduction electrons due to
large Hund's-rule coupling energy\cite{okimoto95}.

Fig. 1 shows the shift in the radio frequency reactance, $dX$ as the
temperature is varied between $300K$ and $70K,$ for single crystals $SC1,SC2$%
, and $SC3.$ . As can be seen from the figure all the three single crystals
show a sharp main transition at $T_{c}=125K$ . Results for the polycrystal
sample $PC$ (not shown) were similar but the transition at $T_{c}$ was
broadened. In the case of sample $SC1$ interestingly, a very strong jump,
representative of a secondary transition, is observed at a temperature $260K$%
, which we call $T^{*}$. It is worth noting that the shift at this
temperature is almost $35\%$ of the shift at $T_{c}$ and has similar
characteristics as that of the transition at $T_{c}=125K$. Such a transition
is not seen in the static dc resistivity and is only observed as a very
small change in slope in magnetization measurements. Samples $SC2$, $SC3$
and $PC$ also show similar but weaker transitions around the same
temperature. While the first transition at $125K$ is the well known $MI$
transition ( which is also observed in the $dc$ $M\left( T\right) $ and $%
\rho \left( T\right) $ measurements), the origin of the transition at $%
T^{*}=260K$ is quite intriguing and worth studying.

In the magnetization measurements, very weak features are observed in the
vicinity of $T^{*}$, but are nowhere as pronounced as in Fig.1. In the
literature similar weak features have been reported in double perovskite but
have not been adequately addressed \cite{moritomo96,bader98,despina98}. In
the present $ac$ measurements, the features are striking and unavoidable. In
this paper we focus on this anomaly - a subsequent paper will discuss the
details of the field and temperature dependence of the rf response over the
entire temperature range. It is important to dwell on these unusual
transitions as they may hold a key to an understanding of the basic
mechanism involved in the $CMR$ materials.

It has earlier been demonstrated from $TEM$ (transmission electron
microscopy) analysis that intergrowths could be present in these materials%
\cite{sesh,bader98}. The intergrowths are missing or extra layers of $SrO$
between $MnO_{6}$ octahedra, which could form higher order $(n>2)$ phases of 
$RP$ series. The unit cell of $327$ may be written as $SrO\left(
La_{1-x}Sr_{x}MnO_{3}\right) _{2}$ , where it is clear that $MnO_{6}$
octahedra are separated from each other by an insulating $SrO$ layer. If the
second transition is thought to be due to impurity $113$ phase which forms
due to missing of $SrO$ layers, then this phase is likely to be $%
La_{0.6}Sr_{0.4}MnO_{3}$. The corresponding transition temperature of this
phase is well above $300K$ \cite{kawano}. Secondly, any phase transition in
the $MnO_{3}$ sytem is, generally, associated with a large CMR at the
transition. From the $\rho \left( T\right) $ it can be inferred that a large
change in magnetoresistance is not observed at $T^{*}$ even though as noted
previously the overall magnetoresistance is substantial. Hence, the
transition at $T^{*}$ that we observe is not due to $La_{0.6}Sr_{0.4}MnO_{3}$
but due to some other mechanism. The transition at $T^{*\text{ }}$could not
be due to any other impurity phases since the experimental $XRD$ \cite
{prellier}and electron diffraction \cite{dechamps} studies show that the
sample is free from impurity phases. Therefore, one can rule out the
intergrowths or impurities playing any role in inducing this transition.

In Fig. 2 we present the effect of external field in the $ab-plane$ $%
(H\,||\,ab)$ and along the $\hat{c}-axis$ $(H\,||\,c).$ As can be seen from
the figure, in the case of $H\,||\,ab$ the change in $dX$ at $MI$ transition
(at $125K)$ decreases when compared to the same in the absence of field and,
interestingly, the transition at $T^{*}$ is almost smeared out. However, in
the case of $H\,||\,c$ the change in $dX$ at $MI$ transition increases and,
the transition at $T^{*}$ disappears as in the case of $H\,||\,ab$. Since
the easy axis of the magnetization is in the $ab-plane$ the decrease /
increase in $dX,$ in the case of $H\,||\,ab$ / $H\,||\,c$, can be understood
in terms of decrease/increase in magnetization, respectively. If the $T^{*}$
is due to an impurity phase (which ought to have an easy-axis in certain
direction), as is expected from intergrowth mechanism, the $dX$ at this
temperature should have opposite responses for $H\,||\,ab$ and $H\,||\,c$
configurations, as in the case of $MI$ transition. The present response,
unequivocally, suggests that the transition at $T^{*}$ is not due to
impurity phases or intergrowths. It instead suggests some kind of ordering
with a weak exchange interaction which gets destroyed in the presence of
moderate magnetic fields.

Recently, Moritomo et al. \cite{moritomo96} have observed a transition
around $275K$ in $x=0.4$ composition. However, no case has been made to
explain it. In the case of $x=0.3$ composition, Kimura et al.\cite{kimura96}
observed a striking secondary $MI$ transition in the $ab-plane$ $\left(
MI_{ab}\right) $ resistivity, in addition to the usual $MI$ transition
around $\ 100K$ . They did not observe such secondary $MI$ transition in the 
$c-axis$ resistivity measurements.. The result was ascribed to the $2D$
ferromagnetic correlations within the $MnO_{2}$ bilayers leading to $MI_{ab}$
transition at $275K$. The $3D$ spin ordering across the nonmagnetic
insulating $(La,Sr)O$ layers was suggested to be taking place at the $T_{c.}$
However, in the present crystal but for the magnitude change in the
resistivities $\rho _{ab}$ and $\rho _{c},$ no further difference is
observed (inset, Fig. 2). In both the cases in the temperature range $%
T_{c}<T<T^{*\text{ }}$ the sample shows insulating behavior.

The $CMR,$ defined as $-\left( \rho \left( H\right) -\rho \left( 0\right)
\right) /\rho \left( 0\right) ,$ is almost $100\%$ for $T_{c}<T<150K$ and
exceeds $10\%$ in the entire temperature range. This $CMR$ behavior is in
sharp contrast with that of the $La_{1-x}A_{x}MnO_{3}$ perovskite, where it
is present only below and close to $T_{c}$ \cite{von,jin,cnrrao,ramirez}.
The enhanced CMR in the present compound could be due to anisotropic
exchange interaction. The intra layer exchange interaction can be understood
in terms of double exchange. Below $T_{c}$ the interlayer coupling is
established which contributes to the $3D$ magnetic ordering. The very large $%
CMR$ observed in the vicinity of $T_{c}$, both below and above, is due to
the alignment of the $t_{2g}$ spins in the presence of applied magnetic
field which reduces the scattering of the carriers by the local spins. The
fact that large $CMR$ is observed even above $T_{c}$ indicates that some
kind of spin correlation is maintained in this region. There could be many
competing interactions/ordering above $T_{c}$ such as antiferromagnetic
superexchange, electron lattice (the Jahn-Teller effect) and charge
ordering. From the inelastic neutron scattering study it has earlier been
inferred, in the case of $x=0.4$ double perovskite, that the spins in the
neighboring layers are strongly canted\cite{osborn98} and that there exists
in-plane antiferromagnetic $(AF)$ correlations above $T_{c}$ which lead to
insulating behavior \cite{perring}. In the present crystals also from the
inelastic neutron scattering study\cite{chat} it has been observed there is
a coexistence of antiferromagnetism and ferromagnetism above $T_{c}$ . Such
a competition between ferromagnetic double exchange and antiferromagnetic
super exchange has frequently been observed in the manganites which shows a
large negative CMR at temperatures above $T_{c}$ because of the field
suppression of the AF fluctuations\cite{kuwahara}.

However, a careful analysis of the present results of $x=0.4$ composition
indicate that the transition at T$^{*}$ is not due to antiferromagnetic
correlations but, probably, due to weak ferromagnetic correlations. The
results show that main transition at $T_{c}$ and the secondary transition at 
$T^{*}$ are similar in nature.

In Fig. 3 the change in reactance, $dX$ in the presence of $0.3T$ magnetic
field, as a function of the angle between field and $\hat{c}-axis$ of the
crystal at various temperatures is shown. The response at 10K can be
considered purely magnetic since the field has no effect on $\rho $ at this
temperature (inset, Fig. 2). It can be seen that the overall response shows
a reduction in permeability with the increase in the angle from 0$^{o}$ to 90%
$^{o}.$ Since the easy-axis of the $327$ phase is in the ab-plane this
reduction in $\mu $ is expected. However, at low angles one can see an
increase in permeability which we identify as the response due to secondary
phase. It is worth noting that the minima are separated by exactly 90$^{o}$.
Therefore, it is not improper to conclude that the easy axis of the
secondary phase is along $\hat{c}-axis$. This result coupled with the $%
dX\left( T\right) $ provides a strong evidence that the transition at $T^{*}$
is ferromagnetic in nature and is independent of the ferromagnetic
transition at $T_{c}$. Above $T_{c}$\ ($T=188,280$ and $300K$) it can be
seen that a much reduced response (change in $\mu $ ) persists. The reason
for this small change is the existence of ferromagnetic fluctuations above $%
T_{c}$ which is evidenced by the ferromagnetic hysteresis loop even at $295K$%
\cite{prellier}. The aforementioned field dependent anisotropy (Fig. 3) and
temperature dependent (Figs. 1\&2) measurements clearly suggest two kinds of
ferromagnetic correlations occurring intrinsically.

\smallskip It is very interesting to note that the neutron diffraction study
by Mitchell et. al\cite{mitchell96} shows a clear change in slope $%
d(a,b-axis)/dT$ at 260K in the a,b-axis vs temperature plot. However, they
do not observe any change in $d(c-axis)/dT$ at 260K. Together, these two
points clearly indicate that the T$^{*}$ is most probably associated with
spin-lattice coupling and the associated ferromagnetic correlations lie in
the ab-plane. Local lattice or Jahn-Teller distortions of the perovskite
unit cells could lead to spin-lattice coupling. The coupling of the double
exchange to the lattice degrees of freedom in manganites is theroreticallly
studeid by Millis et al.\cite{millis} and Roder et al.\cite{roder96} in
terms of magnetoelastic polarons. The deviation from linearity in the
temperature dependence of lattice parameters of $%
La_{0.6}Y_{0.07}Ca_{0.3}MnO_{3}$\cite{ibarra95} and $%
La_{0.65}Ca_{0.35}MnO_{3}$\cite{dai96} has, earlier, been interpreted in
terms of a transition from large to small polarons. In the present case, the
suggested 2D nature of the secondary transition indicates that it is
probably a transition from large to small magnetoelastic polarons.

The present secondary transition is very important in understanding the
microscopic picture of the manganites. Such intrinsic spin-lattice coupling
can be sensitive to impurities in the crystal, which may be the reason for
not observing $T^{*}$ clearly in samples $SC2$ and $SC3$. It is worth
mentioning that similar high frequency studies on very high quality single
crystals of perovskite $YBa_{2}Cu_{3}O_{7}$ high temperature superconductor
(HTS) gave rise to many novel features\cite{sri,zhai} which are not observed
in general. In conclusion we have presented the temperature and field
dependence of dynamic radio frequency permeability study of double
perovskite, $La_{1.2}Sr_{1.8}Mn_{2}O_{7}$. These new results indicate a
secondary transition at $260K$ similar to the main $MI$ transition at $125K.$
This anomalous secondary transition could be due to weak ferromagnetic
correlations which are independent of the ferromagnetic ordering at $T_{c}$,
which affect the dynamic permeability but are weakly coupled to charge
transport.

We thank J.B.Sokoloff for useful comments. Work at Northeastern was
supported by NSF-9711910 and NSF-9623720. This joint collaboration was
supported by a US-France NSF-CNRS grant NSF-INT-9726801.

$^{a}$ electronic address :\ srinivas@neu.edu

\end{multicols}%

\begin{figure}
\caption{Temperature dependence of radio frequency reactance for single crystals SC1, SC2 and
SC3.}
\label{fig1}
\end{figure}%

\begin{figure}
\caption{Temperature dependence of rf reactance in the presence of magnetic field, 
H=0.6T for sample SC1. The change $ dX$ at $T_c$ increases when the field is parallel to the
 c-axis contrary 
the change when the field is parallel to the ab-plane. Inset shows the temperature dependence 
of resistivity in the ab-plane and along c-axis, both in the presence and absence of field.}
\label{fig2}
\end{figure}%

\begin{figure}
\caption{Angular variation of rf reactance in the presence of $0.3T$ field at various
 temperatures for sample SC1. The minimun at $0^{\circ}$ is 
representative of secondary transition while the minimum at $90^{\circ}$  represents
 ferromagnetic ordering below $T_c$.}
\label{fig4}
\end{figure}%

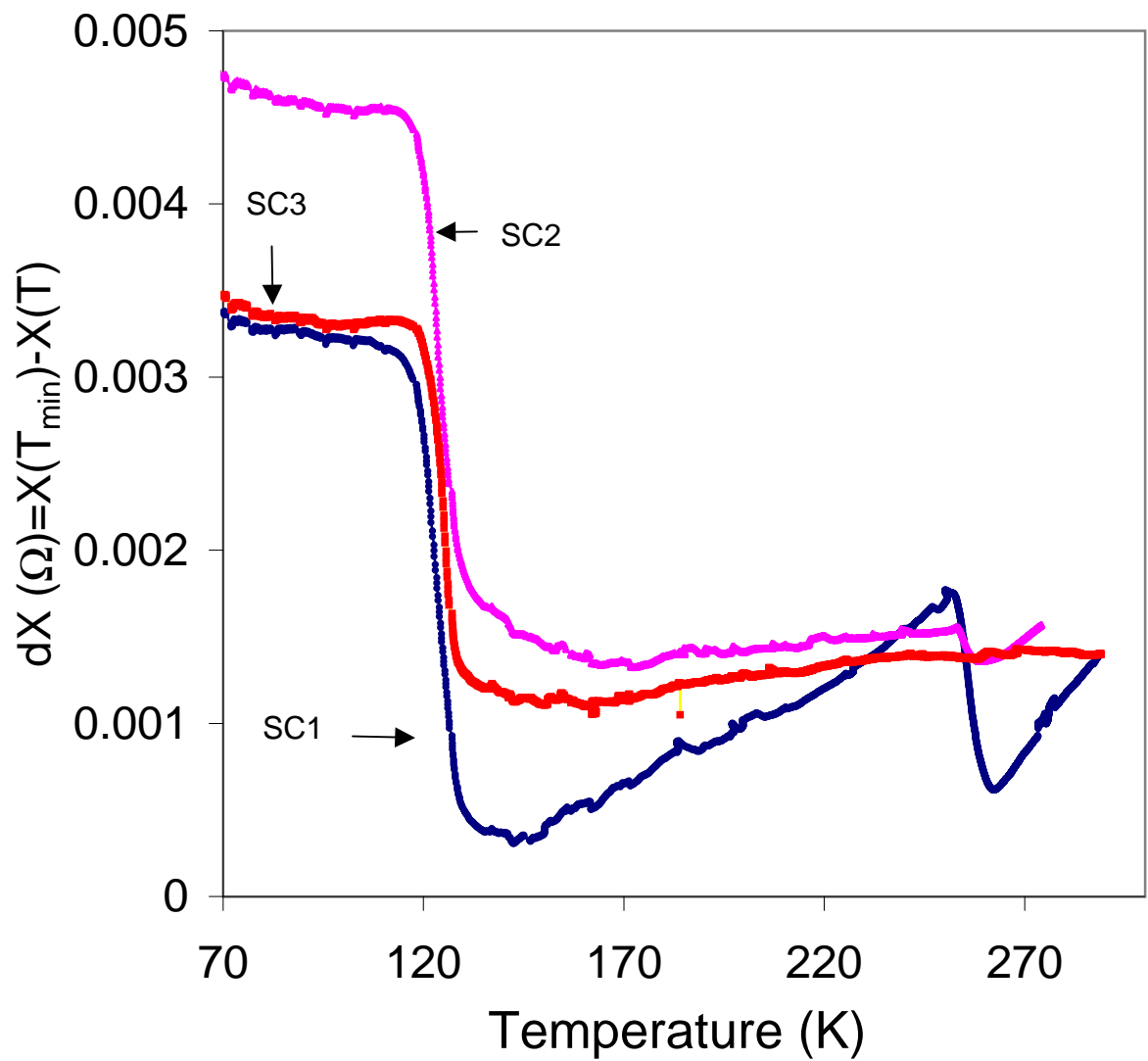
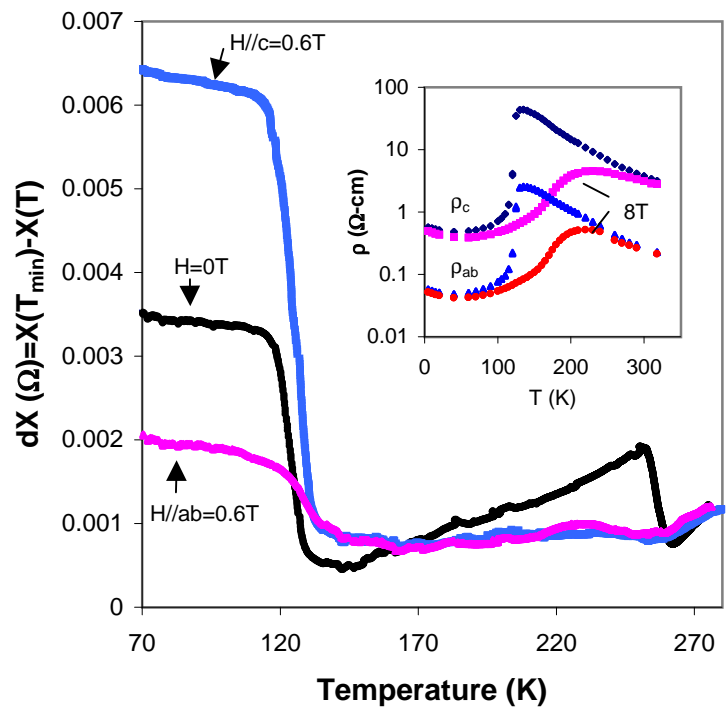
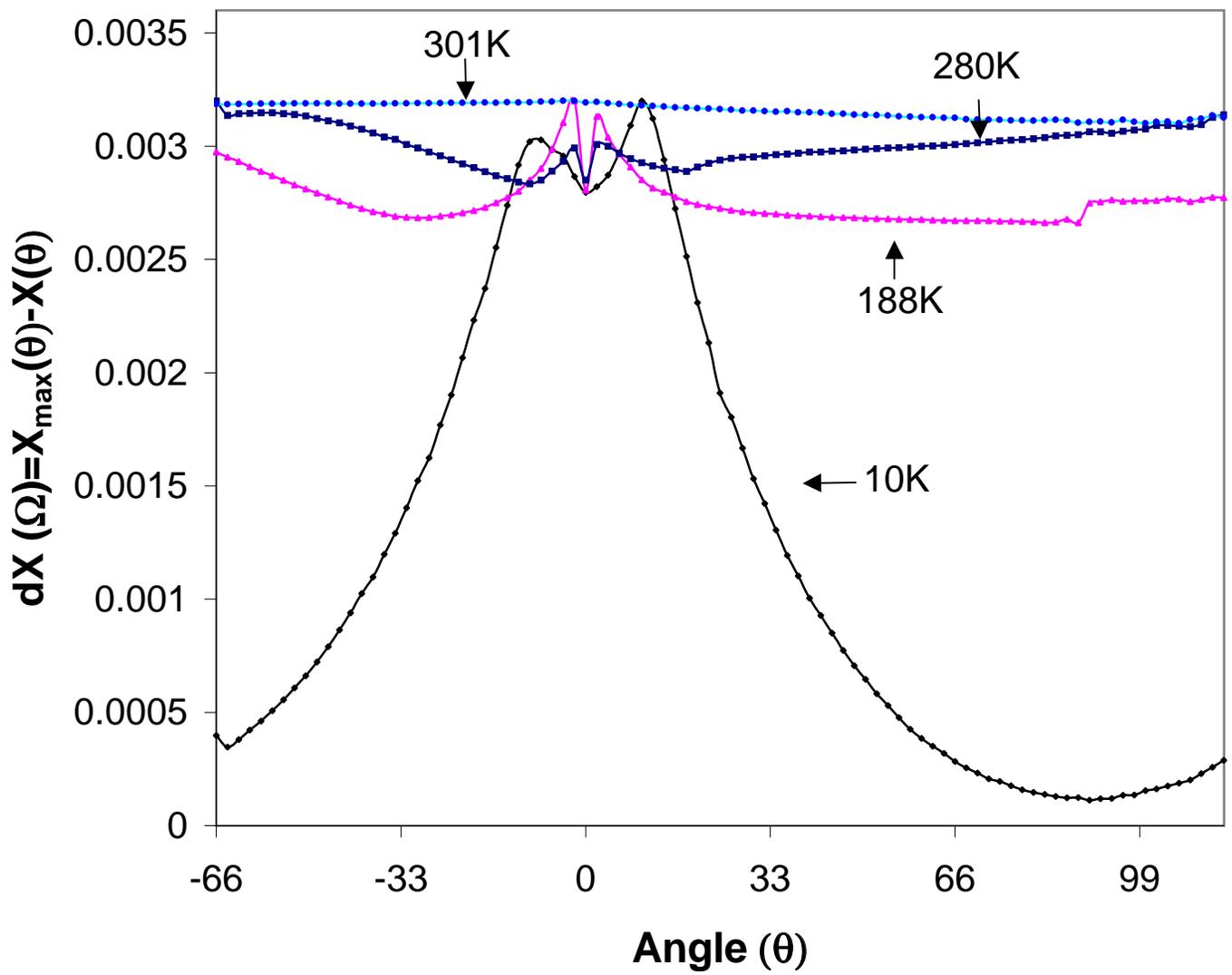
\end{document}